\documentclass[pre,twocolumn,showpacs,amsmath,amssymb,floatfix]{revtex4}

\usepackage{graphicx}
\usepackage{times}

\begin{document}

\newcommand{\Fharm}{{E_{\rm harm}}}
\newcommand{\muu}{\boldsymbol{\mu}}
\newcommand{\Ns}{{N_s}}
\newcommand{\Jaf}{J}   
\def\secc#1{{\it #1} ---}

\def \be {\begin{equation}}
\def \ee {\end{equation}}
\def \bea {\begin{eqnarray}}
\def \eea {\end{eqnarray}}

\def \nn {\hat{\bf n}}
\def \zz {\hat{\bf z}}
\def \hh {{\bf h}}
\def \ss {{\bf s}}
\def \HH {{\cal H}}
\def \II {{\bf I}}
\def \Tr {{\rm  Tr}}
\def \Fharm {{\cal F^{\rm h}}}
\def \MM {{\bf M}}
\def \UU {{\bf U}}
\def \muutot {\muu_{\rm tot}}
\def \LL {{\bf L}}
\def \QQ {{\bf Q}}
\def \la {{\langle}}
\def \ra {{\rangle}}
\def \ising {{\eta}}
\def \iloop {{\tau}}
\def \roo{{r}}
\def \soo{{s}}
\def \noo{{\nu}}
\def \kag {{kagom\'e}}
\def \pyr {pyrochlore}
\def \eqr#1{(\ref{#1})}


\title{Order by disorder and gauge-like degeneracy in 
        quantum pyrochlore antiferromagnet}

\author{Christopher L. Henley}
\affiliation{Laboratory of Atomic and Solid State Physics, Cornell University,
Ithaca, New York, 14853-2501}


\begin{abstract}
The (three-dimensional) pyrochlore lattice antiferromagnet 
with Heisenberg spins of large spin length $S$ is a highly 
frustrated model with an macroscopic degeneracy of classical ground states.
The zero-point energy of (harmonic order) spin wave fluctuations
distinguishes a subset of these states.  I derive an approximate
but illuminating {\it effective Hamiltonian}, acting within the subspace
of Ising spin configurations representing the {\it collinear}
ground states.  It consists of products of Ising spins around 
loops, i.e has the  form of a $Z_2$ lattice gauge theory.
The remaining ground state entropy is still infinite but
not extensive, being $O(L)$ for system size $O(L^3)$.
All these ground states have unit cells bigger than those
considered previously.
\end{abstract}

\pacs{PACS numbers: }

\maketitle

What is the nature of the ground state of a highly
frustrated antiferromagnet with spin length $S$?
For unfrustrated (and simply frustrated, e.g. triangular)
antiferromagnets, a valid recipe is to find the
classical ground state (i.e., solution of mean-field theory)
and dress it with spin-wave fluctuations. 
But many frustrated magnets
with vector spins have a large classical ground state
manifold, having more dimensions than the two or
three guaranteed by spin-space rotational symmetry. 
So, to address frustrated systems with nontrivial ground state
degeneracies, one must furthermore identify how
quantum fluctuations elect one of the classical states
to be the basis of the real ground state (or,  alternatively,
mix them all, producing a spin-disordered ground state).

Consider the magnon zero-point energy~\cite{Shen82} at harmonic order, 
$\Fharm(\{ \nn _i \}) \equiv \sum _m \frac{1}{2} \hbar \omega_m$,  
where $\omega_m$ runs over all modes of spin waves fluctuating
around a particular classical ground state $\ss_i=S\nn_i$ with
unit vectors $\{ \nn _i \}$.
[Note that $\Fharm$ is defined {\it only} on classical ground states:
it is just the first term, after the classical one, 
of an  expansion in $1/S$, and thus dominates in the 
large-$S$ (semiclassical) limit.]
Inequivalent ground states do have different spectra
$\{ \omega _m \}$, so $\Fharm(\{ \nn _i \})$
does take different values, and singles out a particular ground state.
In simpler cases with (non massive)  degeneracies~\cite{HenOD}
-- e.g. face-centered cubic (fcc) antiferromagnets --
$\Fharm(\{\nn_i \})$ is minimized  for a unique (modulo rotations) 
classical configuration $\{ \nn _i \}$: this long-range
ordered state is our answer  (for sufficiently large $S$).
Such breaking of a degeneracy by quantum or thermal fluctuations, 
or quenched disorder, has been called ``order by disorder''~\cite{Vill80,HenOD}.

``Highly frustrated'' magnets have similar but larger degenerate manifolds,
in that the number of independent angle parameters is extensive in the
system size.  The best-known examples are the two-dimensional \kag~ and
the three-dimensional pyrochlore  lattice,
built respectively from corner-sharing triangles or tetrahedra.
My aim here is to ascertain for which classical configuration the
quantum fluctuation energy $\Fharm$ is smallest; if unique
(modulo spin rotation and lattice symmetries),
this configuration would give the long-range order
of the true ground state.

This paper will address only the $T=0$ ordered 
state of the large-$S$, pure Heisenberg quantum antiferromagnet 
with nearest-neighbor couplings on the pyrochlore lattice, which 
might describe the low-$T$ state that ZnCr$_2$O$_4$ ($S=3/2$)
is approaching before it undergoes a structural transition~\cite{Lee02}.
Thus our expectation of long-range order does not contradict 
the evidence for spin-disordered (spin liquid or valence bond crystal)
states at $S=1/2$ \cite{Ha91,Can98}, or the lack of collinear order
as $T\to 0$ in the classical model~\cite{Reim92,Moe98}.

\secc{Effective Hamiltonian idea}
A major barrier to identifying the selected state is
that there is an infinity of states to expand around. 
This problem is often approached by computing and comparing
the $\Fharm$ values for special states 
which have exceptional symmetry~\cite{Chu92} or a small 
magnetic cell~\cite{old-harm}.
For these special ground states, a numerically exact $\Fharm$
can be found by integrating over the magnetic Brillouin zone.
Yet in principle we need to know $\Fharm$ for each of the continuously
infinite ground states, almost all of which are nonperiodic; indeed,
the true minimum might {\it not} lie in this special subspace
[In the present case, it has a larger magnetic cell.]

Instead, my general approach is to derive 
$\Fharm(\{ \nn _i \})$ as 
an {\it effective Hamiltonian}~\cite{Hen01-HFM},
for a {\it generic} classical ground state, 
in a crude approximation that
has no controlled small parameter, but results in an 
elegant form (products of spins around loops). 
This is faithful in the sense that the approximate energies 
are (mostly) in the same order as the exact ones, 
and it displays clearly which attributes of a configuration
affect its harmonic spin-wave energy.

The {\it collinear} states, in the pyrochlore system, 
are the special subset of states in which $\nn_i = \ising_i \nn$
for all $i$ with $\ising_i=\pm 1$.  Apart from a global spin axis,
they are simply the ground states of the {\it Ising}
pyrochlore antiferromagnet~\cite{An56}; this {\it discrete}
subset still has a massive degeneracy, expressed now as 
an extensive ground-state entropy ($\sim N \ln(3/2)$).
I assert (based on many checks, but no proof) that every collinear state
is a {\it local} minimum of $\Fharm$~\cite{FN-collinear},
in accord with the general behavior in exchange-coupled systems~\cite{Shen82,HenOD};
I further conjecture that the optimal $\Fharm$
is attained on a collinear state.
So we may henceforth limit ourselves to collinear states
(an enormous reduction of the set)
and seek the {\it discrete} effective Hamiltonian $\Fharm(\{ \ising_i \})$.

In the \kag~ case the discrete coplanar states~\cite{Ritch93}
play the same role as collinear states do for the pyrochlore; 
there, it turned out that it {\it every} coplanar state has the same spectrum
$\{ \omega _m \}$, hence $\Fharm ()$ was {\it independent}
of the discrete configuration (among coplanar states).
A self-consistent {\it anharmonic}  calculation
was required to reduce this degeneracy to a unique state
with the final conclusion of long-range order in
the  $\sqrt 3 \times \sqrt 3$ state~\cite{Chu92,Chan94}.
Note the independence of the spectrum depended
on the fact that $\nn_i\cdot \nn_j =-1/2$ for neighbor spins 
in every coplanar \kag~ state.

On the other hand, in collinear pyrochlore
states, $\nn_i \cdot \nn _j = \pm 1$, hence 
the classical degeneracy {\it is} split by $\Fharm$.
The goal of this paper is to discover how it is split:
what is the (approximate) analytic form of $\Fharm$,
what is its energy scale,
which spin pattern gives the minimum $\Fharm$,
and how large is the remaining degeneracy.
It will be shown that
there is an infinite (discrete) number of minimum-energy states, but
the ground-state entropy is {\it not} extensive in the system size. 

\secc{Hamiltonian and equation of motion}
The pyrochlore lattice is the medial lattice of a
diamond lattice, i.e. 
the pyrochlore sites are the midpoints of the diamond bonds.
Every tetrahedron of sites in the pyrochlore lattice is centered
by a diamond site.  
With $N$ unit cells, there are $\Ns=4N$ spins.
Our Hamiltonian contains only isotropic, antiferromagnet exchange 
coupling between nearest neighbor spins $\ss_i$ of 
length $S$ on the pyrochlore lattice:
   $\HH= \Jaf \sum _{\la ij \ra} \ss_i \cdot \ss_j$,
where ``$\la ij \ra$'' counts each neighbor pair just once.

Now, define the tetrahedron spin 
$\LL_\alpha \equiv \sum _{i\in \alpha} \ss_i$
where $\alpha$ (and other Greek indices) runs over the
tetrahedron centers, and ``$i\in \alpha$'' means $i$ is one
of the four sites in tetrahedron $\alpha$. 
Then the Hamiltonian can be rewritten 
$\HH = 
\frac {1}{2} \Jaf \sum _\alpha \LL_\alpha^2$
and the classical ground states are the
(very many) states satisfying
$\LL_\alpha =0.$

The $\{\omega_m \}$ which enter $\Fharm$ are the same as the 
eigenfrequencies of 
the (linearized) classical dynamics. 
The classical equations of motion are
    \be
             \hbar \dot \ss_i = S^{-1} \ss_i \times \hh_i = JS^{-1} \ss_i\times
             \sum _{\alpha: i\in \alpha} \LL_\alpha
    \label{eq:spindyn}
    \ee
where 
$\hh_i \equiv {\delta \HH}/{\delta \ss_i}$
$ = J \sum _{\la ij\ra} \ss_j 
=  J \sum _{\alpha: i\in \alpha} (\LL_\alpha-\ss_i)$.
In explaining the slow dynamics of the {\it classical} system,
Ref.~\cite{Moe98} had noted that \eqr{eq:spindyn} implies
    \be
      \hbar \dot{\LL}_\alpha 
        = - S^{-1} \Jaf \sum_\beta \ss_{i(\alpha\beta)}  
      \times \LL_\beta, 
    \label{eq:tetdyn}
    \ee
where the sum is over the four tetrahedra $\beta$ that are nearest
to $\alpha$, and $i(\alpha\beta)$ denotes the (unique) spin shared by
$\alpha$ and $\beta$.  
Without loss of generality, take
${\ss_i}^{(0)} \equiv \ising_i S \zz$
in our (assumed) collinear state,
and parametrize the deviations
as $\delta\LL_\alpha(t) = 
(\delta L_{\alpha x}(t), \delta L_{\alpha y}(t), 0)$:
the linearized spin dynamics reduces to 
  \be
        \delta \dot{\LL}_{\alpha} = 
                       - S \Jaf \sum_\beta \mu_{\alpha\beta} 
                       \zz \times \delta \LL_{\beta}
  \label{eq:tetlin}
  \ee
in which a $2N \times 2N $ matrix $\muu$ is defined with 
elements $\mu_{\alpha \beta} \equiv \ising_{i(\alpha \beta)}$.
So, via the trick of using tetrahedron spins,
the dynamical matrix {\it is} the classical spin configuration;
Now, if we can only massage the formulation so it appears as a perturbation,
the expansion will generate an effective Hamiltonian.

\begin{figure}
\includegraphics[width=6.66cm,angle=0]{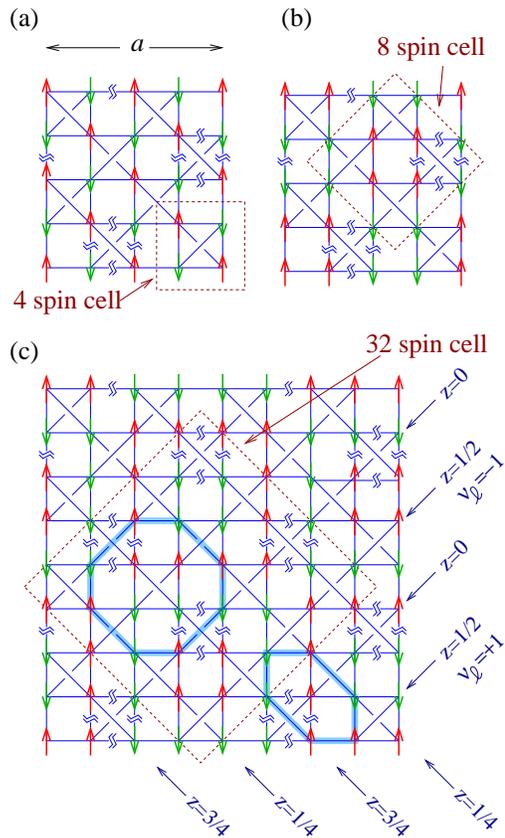}
\caption
{[COLOR ONLINE].
Spin configurations on the pyrochlore lattice.
Sites shown lie in a slice of thickness one unit
cell, normal to the page, and all these configurations
repeat in that direction.  
The hexagon and puckered octagon loops
are superposed (dashed).
The diagonal lines
lie in the plane of the page, forming four families
at different heights $z$ (marked).  Lines running in the $x$ or $y$
directions are tilted $45^\circ$ from the vertical, and
are shown broken where they connect to an adjacent slice.
The lattice unit cell (outlined in (a)) contains  4 spins
and has lattice constant $a$.
Configurations (a) and (b) are ``gauge'' equivalent and
have the highest possible harmonic zero-point energy
$\Fharm$. Configuration (c), 
with its 32-spin magnetic cell outlined, 
is the simplest one of many that have the lowest  $\Fharm$.
[Energies for these configurations are in Table~\ref{tab:Fharms}.]
}
\label{fig:spins}
\end{figure}

\secc{Trace expansion and loop effective Hamiltonian}
Iterating \eqr{eq:tetlin} gives the eigenvalue equation 
   \be
    (\hbar\omega)^2 \delta \LL_\alpha =
          -\hbar^2 \delta \ddot{\LL}_{\alpha} = 
            (SJ)^2 \sum_\gamma (\muu^2)_{\alpha\gamma}\delta\LL_{\gamma} .
   \ee
A square root and trace now express the desired 
effective Hamiltonian~\cite{Hi05b,FN-zeromodes}:
    \be
      \Fharm(\{ \ising_i \} ) 
           \equiv {1\over 2} \sum_m \hbar \omega_m
           \equiv {1\over 2} JS \; \rm{Tr} \left( [\muu^2] ^{1/2}\right).
    \label{eq:trace}
    \ee
The matrix has constant diagonal elements,
$(\muu^2)_{\alpha \alpha} = 4$,
so let us break it up as 
$\muu^2 = 4 \II + \MM$, 
where $\II$ is the identity matrix; $\MM$ has elements
of form $\ising_i\ising_j$ and only connects 
(nearest) even sites of the diamond lattice.  
Next, insert into \eqr{eq:trace}
a formal Taylor expansion of $(4\II +\MM)^{1/2}$
Every term $\Tr(\MM^n)$ is a product 
$\prod_{k=1}^{2n} \ising_{i_k}$
over a closed walk of $2n$ steps on the diamond lattice.
A step retraced twice contributes a trivial factor $\ising_i^2 =1$.
Thus
   \begin{equation}
       \Fharm = {\Fharm}_0- \sum _\Gamma K_\Gamma \prod _{i\in \Gamma} \ising_i
   \label{eq:Heff-loops}
   \end{equation}
where $\Gamma$ runs over loops (without acute angles)
in the pyrochlore lattice.
Eq.~(\ref{eq:Heff-loops})  has exactly the form of an 
{\it Ising lattice gauge model}~\cite{Kogut79}
on a diamond lattice.
The shortest loops $\Gamma$ are 6-step and 8-step loops, which 
appearing first in the  $\Tr(\MM^3)$ and $\Tr(\MM^4)$
terms (respectively) of the expansion.
Such a hexagon and puckered octagon are outlined
in Fig.~\ref{fig:spins}; call their coefficients
$K_6$ and $K_8$.

\begin{table}
\caption{Exact versus approximate zero-point energy, 
as a multiple of $JS N_s$.
The magnetic unit cell has $N_{\rm mag}$ spins.
Last column is from Eq.~\eqr{eq:Heff-comp} with
$A=4$, from the expansion of \eqr{eq:trace}
to $O(\MM^4)$.}
\begin{tabular}{|llll|ll|}
\hline
State   &  $N_{\rm mag}$  &  $\la \iloop_\roo\ra$ &  $\la \iloop_\roo \iloop_\soo\ra$ &
${\Fharm}_{\rm exact} $ & $A=4$   \\
\hline  
a,b  & 4,8 &   +1  &  +1  & 0.4498      & 0.439   \\
c    &  32 &   -1  &  +1  & 0.4245      & 0.414   \\
d    &  64 &    0  &  0   & 0.4460      & 0.430    \\
\hline
\end{tabular}
\label{tab:Fharms}
\end{table}

We can rewrite \eqr{eq:Heff-loops} in a more convenient (for some purposes) form.
Let $\iloop_\roo \equiv \prod_{i\in{\rm hexagon\ }\roo} \ising_i = \pm 1$,
the product around one of the hexagons.  This is a convenient set of
variables, since distinct but ``gauge'' equivalent Ising states map to the
same configuration of $\iloop_\roo$'s, and any product around a longer loop can 
be written in terms of $\iloop_\roo$'s.  We place the value $\iloop_\roo$
at the center of each its hexagon; these sites form 
a new pyrochlore lattice complementary to the old one (and having
the same lattice constant).
In this notation, \eqr{eq:Heff-loops} becomes~\cite{FN-countloops}
   \be  
        \Fharm = -K_6 \sum_\roo \iloop_\roo 
                 - \frac{1}{2} K_8 \sum _{\la\roo,\soo\ra} \iloop_\roo \iloop_\soo + \ldots
   \label{eq:Heff-comp}
   \ee
This has the appearance of an Ising Hamiltonian
with an external field $K_6<0$, and a coupling $K_8>0$;
both terms are satisfied by 
     $\iloop_\roo\equiv -1$, i.e. every hexagon
has a negative loop product.
Configurations of $\{ \ising_i \}$ 
that realize this and satisfy the
classical ground state condition, $\sum_{i\in \alpha} \ising_i =0$
-- which do indeed exist -- are the ground states to harmonic order 
[i.e. to $O(JS)$].
Fig.~\ref{fig:spins}(c) shows the simplest harmonic ground state 
to construct, with 32 spins in the magnetic unit cell; the smallest 
cell has 16 spins.
[All previously studied cases were $Q=0$ states with a 4-spin
magnetic cell, for which the collinear state always has 
$\prod_{i\in \Gamma} \ising_i = +1$, the worst case.]

In Table~\ref{tab:Fharms}, the numerically exact 
zero point energy is compared to (\ref{eq:Heff-loops}), 
taken to order $\MM^4$.
The configurations $\{ \ising _i \}$ were  constructed to have 
different average values (as listed) of the terms in \eqr{eq:Heff-comp}.
Configurations (a,b,c) are shown in Fig.~\ref{fig:spins}.
A fit based on just the three exact energies in 
Table~\ref{tab:Fharms} would give
$K_6^{\rm fit} = -1.3\times 10^{-2}$
and 
$K_8^{\rm fit} = +1.8\times 10^{-2}$.
The $O(\MM^3)$ analytic result is
$K_6^{\rm th} = -3/2^8 = -1.2\times 10^{-2}$, 
and the $O(\MM^4)$ term gives
$K_8^{\rm th} = 15/2^{11} = +0.7\times 10^{-2}$.

It will be seen that a term in \eqr{eq:Heff-loops} 
for a $2n$-step loop actually gets contributions from infinitely many 
terms in the expansion of \eqr{eq:trace},
as terms in $\MM^m$ with $m>n$ can represent
the loop $\Gamma$ plus retraced steps.
In a sense, the higher powers of $\muu$ contain additional contributions
of $\II$; one might  speculate that a different breakup, $\muu^2= 4A\II + \MM'$,
is more accurate.  As the spectrum of $\muu^2$
fills the interval $[0,16]$, 
one choice is to expand around its center, i.e. $4A=8$, 
to optimize the convergence properties;
the original choice $4A=4$ is also natural since $\MM'$ has no diagonal terms.
As an ad-hoc compromise, I suggest the geometric mean $A=\sqrt2$;
exactly that value turned out to be optimal
in a subsequent, systematic graphical resummation, using a Bethe lattice
approximation~\cite{Hi05b}.
[Those numerical energies 
agreed to $\sim$1\% with numerically exact integration
of the zero-point energy for a large sample of classical
ground states~\cite{Hi05b}.]

\secc{Gauge-like symmetry and ground-state degeneracy}
Eq.~\eqr{eq:trace} has an exact ``gauge'' invariance.
Let $\ising' _{i(\alpha,\beta)} = u_\alpha u_\beta \ising _{i(\alpha,\beta)}$
where $u_\alpha = \pm 1$ arbitrarily;
in matrix notation,
$\muu' = \UU \muu \UU^{-1}$, 
where $\UU = {\rm diag}(\{ u_\alpha \})$.
Thus $\muu'$ is similar to $\muu$, the two configurations
have the same harmonic eigenvalue spectrum and exactly the 
same value for $\Fharm ( \{\ising_i \} )$: in other words,
this is a generic way to make $\{ \ising_i' \}$ {\it degenerate}
with  $\{ \ising _i\}$.
I place the term ``gauge'' in quotes, since
genuinely gauge-equivalent configurations are the {\it same} state which
has been labeled redundantly, whereas here they are {\it distinct} 
quantum states.  Note too that we must uphold an independent 
(non-``gauge''-invariant) condition that $\sum _{i\in \alpha} \ising_i=0$
in every tetrahedron, so this spin configuration is still a classical ground state.

We can construct a relatively simple family of ground states
of $\Fharm$.
A slice (as marked in Fig.~\ref{fig:spins})
has sites with four different $z$ values,
constant along the diagonal lines.  Let the $\eta_i$ along
each diagonal line $\ell$ be alternating in sign; to fix an 
overall sign for each, 
choose one $\la 110 \ra$ and one $\la 1 {\bar 1} 0 \ra$ plane normal
to the lines, and let $\noo_\ell$ be the spin in that layer.
Then, the extreme tip sites (as seen in projection) of a hexagon
lie on two lines $\ell,\ell'$ at the same $z$; the remaining
sites are two pairs (lying on lines in the other diagonal direction).
Hence, the loop product for that hexagon is
$\iloop_\soo \equiv \prod_{i\in \soo} \ising_i = \noo _\ell \noo_{\ell'}$.
If we choose $\{ \noo _\ell \}$ to alternate between adjacent lines
in the same $z$-layer, we ensure that $\iloop_\roo\equiv -1$ for
all hexagons as e.g. in Fig.~\ref{fig:spins}(c).
[Configuration (d) in Table~\ref{tab:Fharms} is a layered state, with 
signs $\noo_\ell$ alternating in each layer of spins along $[1\bar{1} 0]$ lines
but all signs $\noo_\ell = +1$ for the lines of spins 
in the $[110]$ direction and hence not a harmonic ground state.]
There is still an arbitrary sign choice in each 
$z$-layer, so this construction gives $2^Z$ distinct ground states, 
where $Z$ is the number of distinct layers (separated by $\Delta z=a/4$
in terms of the cubic lattice constant $a$). This gives a
lower bound of $O(L)$ for the harmonic-ground-state entropy
(in a cube of volume $L^3$), so the entropy 
{\it per spin} is zero in the thermodynamic limit.

We do not yet understand the full set of ground states;
note that ``gauge'' equivalent states exist that are
{\it not} given by the layer construction~\cite{Hi05b}, 
and the total entropy was proven to have an upper bound~\cite{Hi05b}
of $O(Z \ln Z)$, only logarithmically larger than the lower bound.

\secc{Discussion}
In summary, I have computed the $O(JS)$ spin energy term 
which (partly) breaks a classical degeneracy in the 
large-$S$ pyrochlore antiferromagnet with purely
isotropic interactions.
Various complex orderings are found in real pyrochlore
antiferromagnets,  most often explained by elastic distortions or 
dipolar interactions ~\cite{expts}
or -- less interestingly -- by nonnegligible second-neighbor exchange.
It was nevertheless valuable to isolate the role of 
quantum fluctuations here, since materials may be found in which
those perturbations are very small~\cite{MPYNN}, or frustrated
Heisenberg models may be cleanly realized by cold gases in 
optical traps~\cite{San04}.

Two tricks -- the Moessner-Chalker
equation of motion,  Eq.~\eqr{eq:tetdyn} and writing the zero-point
energy as the trace of a matrix, Eq.~\eqr{eq:trace}
-- enabled an 
(uncontrolled) expansion of the $O(JS)$ term of
the energy to give the effective Hamiltonian $\Fharm$.
The result \eqr{eq:Heff-loops}
is a sum of products of Ising spins
around loops.  The problem has an {\it exact} ``gauge'' degeneracy 
implying a ground state entropy, to harmonic order,
of (at least) $O(L)$.
To fully resolve the degeneracy, 
as in the \kag~ case~\cite{Chu92, Chan94},
anharmonic spin-wave calculations are in progress~\cite{Hi05c}.

It was plausible that the quantum system shows greater
order than the classical one;
what is surprising is that the pyrochlore is {\it less}
ordered than the \kag~ Heisenberg antiferromagnet 
in the classical $T\to 0$ case~\cite{Moe98},
but {\it more} ordered in the quantum large-$S$ case, 
since harmonic fluctuations
remove an extensive degeneracy in the pyrochlore but not the \kag~ case. 

The loop effective Hamiltonian
has already been applied to other systems~\cite{Tch-Heff}.
A similar approach made it possible to determine the ground state
within the large-$N$, large-$S$ approximation of the pyrochlore~\cite{Hi05a}.
A more complete analytic derivation~\cite{Hi05b} of \eqr{eq:Heff-loops}
matches numerical fits to $\sim 1$\%; it also highlights the modes (fewer
than in the \kag~ case) which have divergent fluctuations in the
harmonic approximation.

An effective Hamiltonian such as \eqr{eq:Heff-loops}
has value beyond the possibility
(as here) that it leads us to an unexpected ground state.
It can also be plugged in to define
a Boltzmann distribution, such as $\exp(-\beta \Fharm)$, 
which at low but nonzero temperatures is more valid than 
the classical spin ensemble~\cite{Hen01-HFM}.
It also a gives basis on which to build 
more complete or more realistic models, by the
addition of anisotropies, quantum tunneling~\cite{vD93}, 
or dilution~\cite{Shen93,Hen01-HFM}.
Notice how the collinear selection has provided a different 
route than ``spin ice'' to realize an effective Ising model 
in a pyrochlore system;
furthermore, tunneling between collinear states~\cite{vD93} might
realize ring exchanges~\cite{Her04}
(and the consequent $U(1)$ spin liquid).

\secc{Acknowledgments}
I thank M. Kvale,  E. P. Chan, R. Moessner, B. Canals, 
H.~Tsunetsugu, O. Tchernyshyov, P.~Sharma, and especially U. Hizi for discussions.
This work was supported by NSF Grant No. DMR-0240953.


\begin{thebibliography}{99}
\bibitem{FN-collinear} 
I note that $\sum \omega_m^2 \propto {\rm Tr}(\muu^2)$, 
where $\muu$ is a matrix defined later in the paper.
This sum is quickly shown to be {\it independent} of which 
ground state $\{ \nn_i \}$ is (collinear or not).  
But $\sum \omega^4$ attains (the same) maximum for all collinear states, 
and is reduced by any deviations of $\{ \nn_i \}$, 
consonant with the arguments in Ref.~\onlinecite{HenOD}
that collinearity generically maximizes the spectrum's dispersion.
That makes it plausible that $\Fharm \propto \sum |\omega|$ 
has a {\it minimum}.
\bibitem{FN-zeromodes}
The $2N$ deviations $\{ \delta \LL_\alpha \}$ are linear combinations of 
the $4N$ deviations $\{\delta\ss_i \}$, so every eigenfrequency $\omega_m$ 
of the former is a spinwave frequency of the latter. Furthermore, 
it can be shown (Ref.~\onlinecite{Hi05b}) that every mode lost in
the projection is a zero-frequency mode that does not affect $\Fharm$.
\end{thebibliography}
\end{document}